\documentclass[conference]{IEEEtran}
\IEEEoverridecommandlockouts
\usepackage{amsthm} 
\usepackage{tikz}
\newcommand*\circled[1]{\tikz[baseline=(char.base)]{
		\node[shape=circle,draw,inner sep=1pt] (char) {#1};}}
\usepackage{url}
\usepackage{verbatim} 
\usepackage{caption}
\usepackage{subcaption}
\usepackage{xcolor}
\usepackage{algorithmic}
\usepackage[linesnumbered,ruled,vlined]{algorithm2e}
\usepackage{amsmath,amssymb,amsfonts}
\usepackage{relsize}
\usepackage{graphicx}
\usepackage{textcomp}
\usepackage{multirow}
\usepackage{titlesec}
\usepackage{diagbox}
\usepackage[hidelinks]{hyperref}
\usepackage{float}

\usepackage{cite}
\usepackage{diagbox}
\usepackage{amsmath,amssymb,amsfonts}
\usepackage{algorithmic}
\usepackage[linesnumbered,ruled,vlined]{algorithm2e}
\usepackage{graphicx}
\usepackage{multirow}
\graphicspath{{./figures/}}
\usepackage{textcomp}
\usepackage{xcolor}
\def\BibTeX{{\rm B\kern-.05em{\sc i\kern-.025em b}\kern-.08em
		T\kern-.1667em\lower.7ex\hbox{E}\kern-.125emX}}
\DeclareRobustCommand*{\IEEEauthorrefmark}[1]{%
	\raisebox{0pt}[0pt][0pt]{\textsuperscript{\footnotesize #1}}%
}

\begin{document}

\title{In-Network Accumulation: Extending the Role of NoC for DNN Acceleration}

\author{\IEEEauthorblockN{Binayak Tiwari\IEEEauthorrefmark{1}, Mei Yang\IEEEauthorrefmark{1}, Xiaohang Wang\IEEEauthorrefmark{2}, Yingtao Jiang\IEEEauthorrefmark{1}}
	\IEEEauthorblockA{\IEEEauthorrefmark{1}\textit{Department of Electrical and Computer Engineering, University of Nevada}, Las Vegas, USA\\ \IEEEauthorrefmark{2}\textit{School of Software Engineering, South China University of Technology}, Guangzhou, China\\
		Email: \IEEEauthorrefmark{1}btiwari@unlv.nevada.edu, \IEEEauthorrefmark{1}\{mei.yang, yingtao.jiang\}@unlv.edu, \IEEEauthorrefmark{2}xiaohangwang@scut.edu.cn}
}

\maketitle

\begin{abstract}
	Network-on-Chip (NoC) plays a significant role in the performance of a DNN accelerator. The scalability and modular design property of the NoC help in improving the performance of a DNN execution by providing flexibility in running different kinds of workloads. Data movement in a DNN workload is still a challenging task for DNN accelerators and hence a novel approach is required. In this paper, we propose the In-Network Accumulation (INA) method to further accelerate a DNN workload execution on a many-core spatial DNN accelerator for the Weight Stationary (WS) dataflow model. The INA method expands the router's function to support partial sum accumulation. This method avoids the overhead of injecting and ejecting an incoming partial sum to the local processing element. The simulation results on AlexNet, ResNet-50, and VGG-16 workloads show that the proposed INA method achieves $1.22 \times$ improvement in latency and $2.16 \times$ improvement in power consumption on the WS dataflow model.
\end{abstract}


\section{INTRODUCTION}\label{introduction}
The increasing popularity of Deep Neural Networks (DNN) can be attributed to the availability of training data, advancements in semiconductor technology, and rapid evolution in DNN algorithms. However, there is also a disparity between the rate at which DNN architecture is evolving and the underlying hardware that executes DNN to satisfy the need of real-world applications. DNN execution demands high computing and power budget, however, many AI applications demand the execution on hardware with limited computation and power budget. Hence, traditional general-purpose processors are no longer able to cater to this need, which drives the research on domain-specific processors i.e., accelerators \cite{dsa}.
\par
Spatial architecture \cite{tpu} is a popular class of domain-specific architecture where a number of Processing Elements (PEs) are connected via a communication fabric to exploit the computing and communication parallelism. DNN accelerators typically consist of an array of PEs each configured with its own local memory, and control unit, as well as a global buffer that hides the DRAM access latency. These PEs are typically connected via the Network-on-Chip (NoC) \cite{noc} which serves a critical role in the overall performance of the DNN execution \cite{NoCinDNN}. NoC enables parallelism in computation and communication by allowing multiple PE support and parallel data movement. 
\par
The dataflow model determines the processing order and where data is stored and reused, i.e., the way data (including inputs, weights, and partial sums) communication happens between the PEs and the global memory. DNN algorithms can be mapped onto the PEs using various dataflow models \cite{hardwareforML} like Output Stationary (OS), Weight Stationary (WS), Row Stationary (RS), etc. each of them has its own memory usage and energy advantage.
\par
Authors in \cite{minimizingComputation} demonstrated that more than $90\%$ of a certain class of  DNN workload execution time on a single-threaded CPU was spent on a convolution layer. Similarly, in most commonly used DNNs the multiply and accumulate (MAC) operations in CONV and FC layers consume more than $90\%$ of the total operations involved. Due to the volume of MAC operations and data (weights, inputs) involved, the DNN execution bottleneck in hardware can be broadly categorized into communication and computation.
\par
The computation bottleneck is fairly straightforward to resolve. This involves adding or increasing the number of PEs or computation nodes. However, it is also to note that increasing computation resources will increase a fair amount of complexity in the communication infrastructure. Additionally, DNN comes in a variety of shape and sizes which include the diversity in layers, kernels, etc. In order to handle these diverse workloads, the communication needs should be reconfigurable. In this paper, we propose the In-Network Accumulation (INA) method which focuses on reducing the network load and potentially memory transactions that will help in alleviating the communication bottleneck.
\par
The rest of the paper is organized as follows. Section \ref{motivation-ina} explains the related background and motivation behind this study. Section \ref{proposed-ina} presents the proposed architectural support for INA. Section \ref{performance-ina} presents the result of the performance evaluation and finally, Section \ref{conclusion-ina} summarizes this paper.

\section{Background and Motivation} \label{motivation-ina}
\begin{figure}
	\centering
	\includegraphics[width=0.9\linewidth]{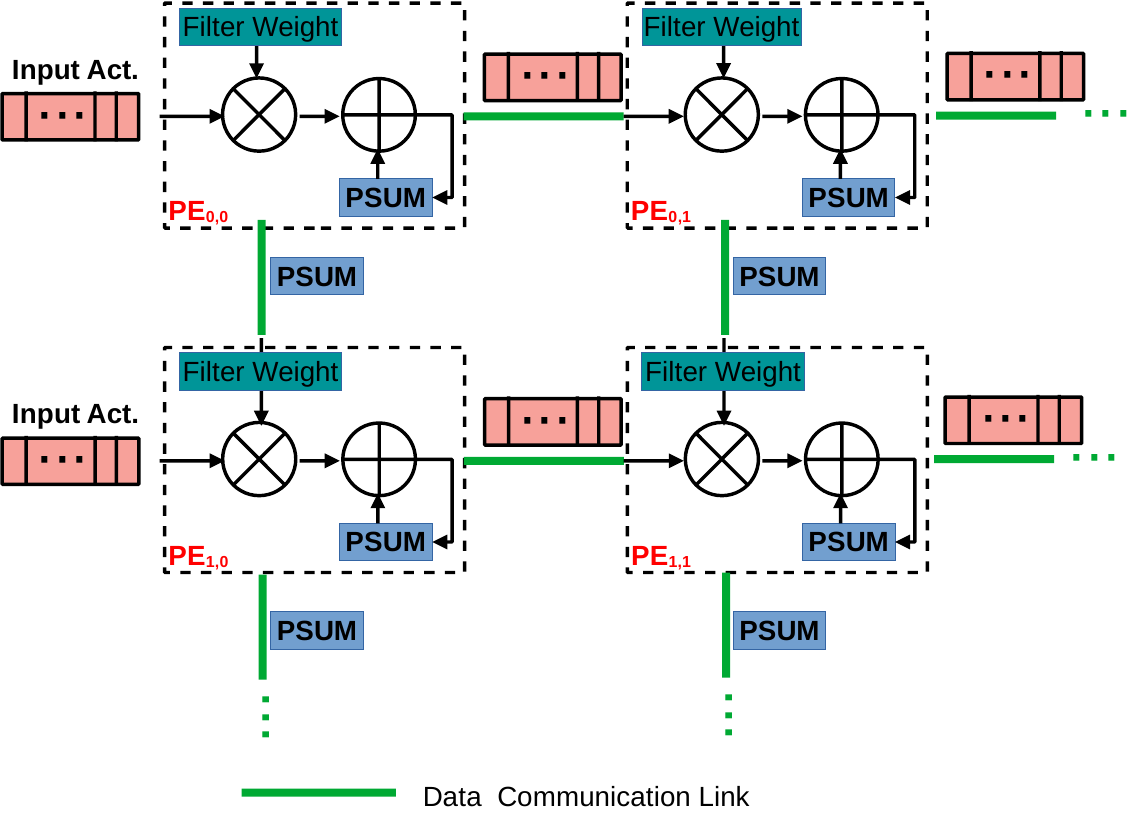}
	\caption{WS dataflow model} 
	\label{fig-ws-df}
\end{figure}
Fig. \ref{fig-ws-df} shows an example of the Weight Stationary (WS) dataflow model where each PE keeps the filter weight stationary in a PE while the input activations and partial sums move across the PE array. When these weights are no longer needed, they will be replaced with other weights.  This setting reuses the weights and minimizes the energy consumption in reading the weights. Google's TPU \cite{tpu} is one example of a design that uses the WS dataflow model.
\par
The dominant computing architecture today relies on memory to provide data to the PE when required. This approach of computing has hit the memory wall i.e., there is a big gap to fill in between the memory latency and CPU execution latency. To overcome this gap, near memory computing or in-memory computing \cite{pim} is considered a promising candidate. The main philosophy behind these architectures is to process the data close to where it is stored. Inspired by this philosophy, in this paper we present in-network accumulation to accelerate the DNN workload in accelerators. The fundamental principle of this architecture is to process the data in the network while routing the data.
\par
There exist enough operations in the DNN workload that can be offloaded to the memory as near-memory computing. Authors in \cite{pim-nn} proposed an in-memory computing architecture for neural network applications, however, this method is not suitable for CNN workloads. This architecture does not offer flexibility in data reuse which leads to frequent reloading of input feature maps and weights during the DNN execution. Similarly, authors in \cite{com} propose a Computing-On-the-Move (COM) architecture to address the shortcomings of \cite{pim-nn}. COM \cite{com} is implemented on a 2D Mesh NoC to enable the inter-memory computation like psum addition.
\begin{figure}
	\centering
	\medskip
	\includegraphics[width=1\linewidth]{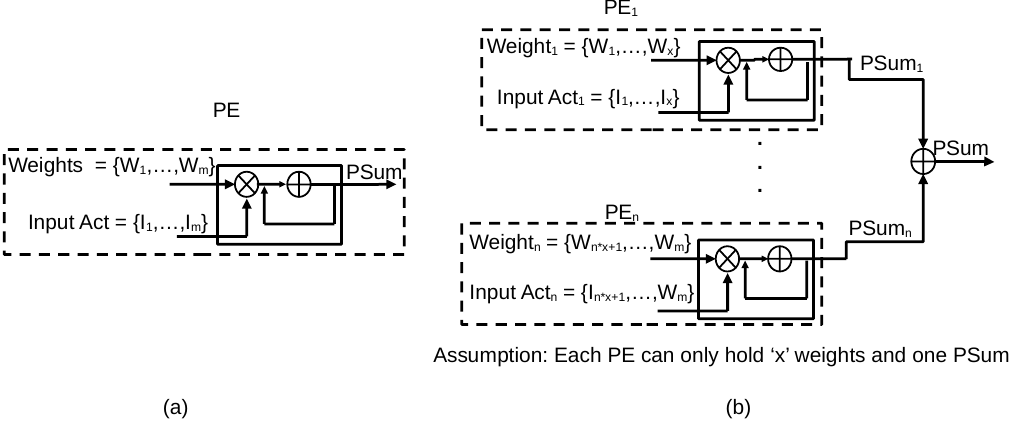}
	\caption{Accumulation Organization (a) OS Dataflow Model (b) WS/RS Dataflow Model}
	\label{fig-ina_org}
\end{figure}

However, the aforementioned architectures cannot be easily integrated with existing accelerator designs adopting different dataflow models. Let us assume a WS dataflow model, where filter weights are stored at local PEs and kept there until all the MAC operations involving the weights are exhausted. The major bottleneck in this approach is to keep the filter weights in the local PE where we have limited scratch memory or a register file. To address this problem, only a portion of the weights are kept in a PE, and overall multiple PEs share and store the weights of one filter. This distribution of filter weights results in the distribution of the partial sum across multiple PEs which need accumulation. A similar distribution also occurs in the RS dataflow model.
\par
Fig. \ref{fig-ina_org} shows the partial sum (psum) generation method which is explained by showing the MAC operation in different dataflow models. A weight vector consisting of $m$ elements ${W_1,...,W_m}$ and an input vector consisting of $m$ elements ${I_1,...,I_m}$ are streamed from the streaming bus to the PE in the OS dataflow model where the psum accumulation is performed at a single PE as shown in Fig. \ref{fig-ina_org} (a). However, due to the distributed nature of weights/inputs in the WS/RS dataflow model particularly due to the memory limitation, psum accumulation needs to happen across different PEs as shown in Fig. \ref{fig-ina_org} (b). Assuming that each PE can hold only $x$ elements from the weights and input vector, only a part of the psum is generated at each PE. This provides an opportunity to optimize the way psum accumulation happens in the WS/RS dataflow model.
\par
In this paper, INA is provided as a solution to perform the psum accumulation on memory constraint models like WS/RS where only a portion of calculation happens in a PE. INA eliminates the additional network traffic due to the limited memory of a PE. The proposed method helps in reducing unnecessary movement of the psum by allowing the psum accumulation perform at the router.

\section{Architectural Support}   \label{proposed-ina} 
\begin{figure}
	\centering
	\includegraphics[width=1\linewidth]{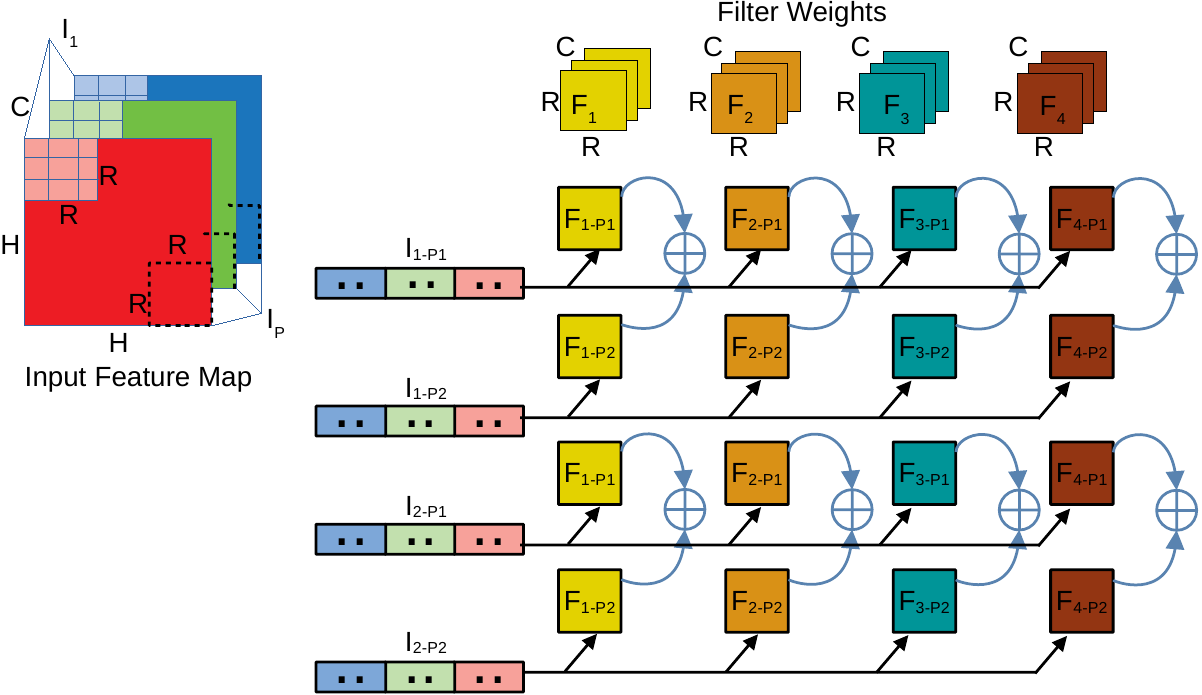}
	\caption{WS Dataflow in 4x4 Mesh NoC} 
	\label{fig-ws_final}
\end{figure} 
The INA design modifies the router to support psum accumulation which makes it easy to integrate with the existing accelerator design. The INA is controlled by a central controller, hence the accelerator can be optimized for individual DNN layers unlike in COM \cite{com} where the dataflow control is distributed. Fig. \ref{fig-ws_final} shows an example of the WS dataflow model on a $4\times4$ mesh. Assume that a CONV operation with a filter in $ \{F_1, F_2, F_3, F_4\}$, input activation in $ \{I_1, I_2,...,I_P\}$ each with a dimension of $C \times R \times R$ is implemented on a $4 \times 4$ mesh where each PE can hold half the elements of a filter. In order to generate one output activation, two PEs are required as shown in Fig. \ref{fig-ws_final} where the filter weights are divided into two parts $P1,\ P2$ and distributed among two adjacent PEs in the same column. After both PEs generate a psum, they can be accumulated across the PEs to get the final output activations. 
\par
\begin{figure}
	\centering
	\includegraphics[width=0.8\linewidth]{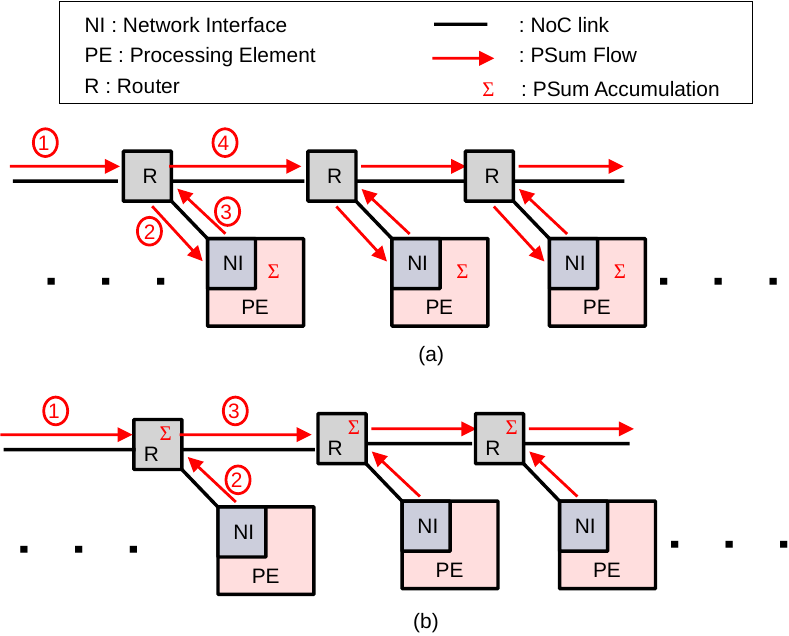}
	\caption{Partial Sum (PSum) accumulation flow for WS/RS dataflow (a) without INA (b) with INA}
	\label{fig-in_network}
\end{figure}
Fig. \ref{fig-in_network} shows the flow for psum accumulation with and without in-network support. Fig. \ref{fig-in_network} (a) shows the flow where \textcolor{black}{ \circled{1}} - \textcolor{black}{\circled{4}} are the sequential steps that need to be performed in order to process the accumulation of the partial sum generated at each node on a WS dataflow model as shown in Fig. \ref{fig-ws-df}. In the absence of in-network accumulation, the incoming packet  \textcolor{black}{ \circled{1}} with partial sum  is first ejected \textcolor{black}{ \circled{2}} to the PE where accumulation happens locally and then the packet with the new accumulation result needs to be injected \textcolor{black}{ \circled{3}} back to the network for the next hop \textcolor{black}{ \circled{4}}.
\par
Fig. \ref{fig-in_network} (b) shows the alternative approach of accumulation of the partial sum generated at each node on a WS dataflow model in the presence of an in-network accumulation unit. Steps \textcolor{black}{ \circled{1}} - \textcolor{black}{\circled{3}} show the flow of partial sum accumulation in the presence of in-network accumulation support. This method of accumulation brings the computation close to the data source and helps in removing the redundant and obvious network transactions. The incoming packet \textcolor{black}{ \circled{1}} can be directly accumulated in the router with the partial sum as generated by the local node \textcolor{black}{ \circled{2}} before forwarding it to the next hop \textcolor{black}{ \circled{3}}. One thing to notice here is ejection and injection of the packets are saved with the support of in-network accumulation which not only helps in improving the latency but also saves the network power consumption.

\subsection{INA Modeling}
INA is beneficial for the WS and RS dataflow models where certain parameters are stored at the PE until reused. Consider a WS dataflow model on a $N \times N$ mesh with 1 PE/router with a memory capacity of $M$ bits. For any CONV layer with $R \times R$ kernel size, $C$ channels, $F$ filters, $O \times O$ output feature map, $q$-bit precision. We can model the INA of partial sums using a series of equations shown below:

Condition to perform INA, iff Equation (\ref{eq6_condition}) holds true.
\begin{equation} \label{eq6_condition}
(C \times R \times R \times q) > M \iff INA
\end{equation}
Number of PEs ($P_\#$) to distribute the filter weights is:
\begin{equation} \label{eq6_numPE}
P_\# = \Big \lceil \frac{C \times R \times R \times q}{M} \Big \rceil
\end{equation}
Rounds of INA ($INA_\#$) to complete one CONV layer on a given $N \times N$ mesh:
\begin{equation} \label{eq6_roundsINA}
INA_\# =  \Big \lceil \frac{F}{N} \cdot \frac{O \times O}{\lfloor \frac{N}{P_\#} \rfloor} \Big \rceil
\end{equation}

Equation \ref{eq6_condition} shows that the condition to perform an INA during a CONV layer execution is dependent on the size of memory for PEs. With sufficient memory there is no need for the INA, however, this is not feasible practically. Hence, to keep $M$ in the practical range, certain dataflow models like WS, RS should distribute the weights, inputs, etc. among multiple PEs leading to the INA as an effective solution. Note that $INA_\#$ represents the total rounds of psum accumulation on a $N \times N$ mesh for a given CONV layer. This means the total number of accumulations in each CONV layer is different. Tables \ref{table-INAAlexnet} and \ref{table-INAVgg} show the number of rounds of INA operation for $8\times8$ and $16\times16$ mesh, respectively. It is seen that there exist enough INA operations that can be optimized. Also, we can see that VGG-16 \cite{vgg16} has a lot of INA rounds compared to AlexNet \cite{alexnet} since VGG-16 has a lot of filters with larger channels which makes the parameter fit in one PE's $M-bits$ memory difficult. Hence, the weights are distributed among different PEs leading to an increase in INA rounds.

\begin{table}
	\vspace{0.45cm}
	\caption{INA Evaluation for AlexNet \cite{alexnet}}
	\begin{center}
		\resizebox{0.95\columnwidth}{!}{
		\begin{tabular}{llllllll}
			\hline
			\multicolumn{1}{|l|}{Layer} & \multicolumn{1}{l|}{R} & \multicolumn{1}{l|}{C}  & \multicolumn{1}{l|}{F}  & \multicolumn{1}{l|}{O} & \multicolumn{1}{l|}{$P_\#$} & \multicolumn{1}{l|}{$INA_\#$, N=8} & \multicolumn{1}{l|}{$INA_\#$, N=16} \\ \hline
			\multicolumn{1}{|l|}{CONV1} & \multicolumn{1}{l|}{11} & \multicolumn{1}{l|}{3}  & \multicolumn{1}{l|}{64} & \multicolumn{1}{l|}{55} & \multicolumn{1}{l|}{1}   & \multicolumn{1}{l|}{NA}   & \multicolumn{1}{l|}{NA} \\ \hline
			\multicolumn{1}{|l|}{CONV2} & \multicolumn{1}{l|}{5} & \multicolumn{1}{l|}{64} & \multicolumn{1}{l|}{192} & \multicolumn{1}{l|}{27} & \multicolumn{1}{l|}{2}   & \multicolumn{1}{l|}{4374} & \multicolumn{1}{l|}{1094} \\ \hline
			\multicolumn{1}{|l|}{CONV3} & \multicolumn{1}{l|}{3} & \multicolumn{1}{l|}{192} & \multicolumn{1}{l|}{384} & \multicolumn{1}{l|}{13} & \multicolumn{1}{l|}{2}   & \multicolumn{1}{l|}{2028}  & \multicolumn{1}{l|}{507}  \\ \hline
			\multicolumn{1}{|l|}{CONV4} & \multicolumn{1}{l|}{3} & \multicolumn{1}{l|}{384} & \multicolumn{1}{l|}{256} & \multicolumn{1}{l|}{13} & \multicolumn{1}{l|}{4}   & \multicolumn{1}{l|}{2704} & \multicolumn{1}{l|}{676}  \\ \hline
			\multicolumn{1}{|l|}{CONV5} & \multicolumn{1}{l|}{3} & \multicolumn{1}{l|}{256} & \multicolumn{1}{l|}{256} & \multicolumn{1}{l|}{13} & \multicolumn{1}{l|}{3}   & \multicolumn{1}{l|}{2704} & \multicolumn{1}{l|}{541}   \\ \hline
			\multicolumn{8}{l}{	*Note: q=32bit, M=32KB, 1 PE/Router}
		\end{tabular}}
	\end{center}
	\label{table-INAAlexnet}
\end{table}

\begin{table}
	\caption{INA Evaluation for VGG-16 \cite{vgg16}}
	\begin{center}
		\resizebox{0.95\columnwidth}{!}{
		\begin{tabular}{llllllll}
			\hline
			\multicolumn{1}{|l|}{Layer} & \multicolumn{1}{l|}{R} & \multicolumn{1}{l|}{C}  & \multicolumn{1}{l|}{F}  & \multicolumn{1}{l|}{O} & \multicolumn{1}{l|}{$P_\#$} & \multicolumn{1}{l|}{$INA_\#$, N=8} & \multicolumn{1}{l|}{$INA_\#$, N=16} \\ \hline
			\multicolumn{1}{|l|}{CONV1} & \multicolumn{1}{l|}{3} & \multicolumn{1}{l|}{3} & \multicolumn{1}{l|}{64}  & \multicolumn{1}{l|}{224} & \multicolumn{1}{l|}{1}  & \multicolumn{1}{l|}{NA} & \multicolumn{1}{l|}{NA} \\ \hline
			\multicolumn{1}{|l|}{CONV2} & \multicolumn{1}{l|}{3} & \multicolumn{1}{l|}{64} & \multicolumn{1}{l|}{64} & \multicolumn{1}{l|}{224} & \multicolumn{1}{l|}{1}  & \multicolumn{1}{l|}{NA} & \multicolumn{1}{l|}{NA} \\ \hline
			\multicolumn{1}{|l|}{CONV3} & \multicolumn{1}{l|}{3} & \multicolumn{1}{l|}{64} & \multicolumn{1}{l|}{128} & \multicolumn{1}{l|}{112} & \multicolumn{1}{l|}{1}  & \multicolumn{1}{l|}{25088}  & \multicolumn{1}{l|}{6272} \\ \hline
			\multicolumn{1}{|l|}{CONV4} & \multicolumn{1}{l|}{3} & \multicolumn{1}{l|}{128} & \multicolumn{1}{l|}{128} & \multicolumn{1}{l|}{112} & \multicolumn{1}{l|}{2}  & \multicolumn{1}{l|}{50176} & \multicolumn{1}{l|}{12544}  \\ \hline
			\multicolumn{1}{|l|}{CONV5} & \multicolumn{1}{l|}{3} & \multicolumn{1}{l|}{128} & \multicolumn{1}{l|}{256} & \multicolumn{1}{l|}{56} & \multicolumn{1}{l|}{2}  & \multicolumn{1}{l|}{25088}  & \multicolumn{1}{l|}{6272} \\ \hline
			\multicolumn{1}{|l|}{CONV6} & \multicolumn{1}{l|}{3} & \multicolumn{1}{l|}{256} & \multicolumn{1}{l|}{256} & \multicolumn{1}{l|}{56} & \multicolumn{1}{l|}{3}  & \multicolumn{1}{l|}{50176}  & \multicolumn{1}{l|}{10036}  \\ \hline
			\multicolumn{1}{|l|}{CONV7} & \multicolumn{1}{l|}{3} & \multicolumn{1}{l|}{256} & \multicolumn{1}{l|}{256} & \multicolumn{1}{l|}{56} & \multicolumn{1}{l|}{3}  & \multicolumn{1}{l|}{50176}  & \multicolumn{1}{l|}{10036}  \\ \hline
			\multicolumn{1}{|l|}{CONV8} & \multicolumn{1}{l|}{3} & \multicolumn{1}{l|}{256} & \multicolumn{1}{l|}{512} & \multicolumn{1}{l|}{28} & \multicolumn{1}{l|}{3}  & \multicolumn{1}{l|}{25088}  & \multicolumn{1}{l|}{5018} \\ \hline
			\multicolumn{1}{|l|}{CONV9} & \multicolumn{1}{l|}{3} & \multicolumn{1}{l|}{512} & \multicolumn{1}{l|}{512} & \multicolumn{1}{l|}{28} & \multicolumn{1}{l|}{5}  & \multicolumn{1}{l|}{50176}  & \multicolumn{1}{l|}{8363} \\ \hline
			\multicolumn{1}{|l|}{CONV10} & \multicolumn{1}{l|}{3} & \multicolumn{1}{l|}{512} & \multicolumn{1}{l|}{512} & \multicolumn{1}{l|}{28} & \multicolumn{1}{l|}{5}  & \multicolumn{1}{l|}{50176} & \multicolumn{1}{l|}{8363} \\ \hline
			\multicolumn{1}{|l|}{CONV11} & \multicolumn{1}{l|}{3} & \multicolumn{1}{l|}{512} & \multicolumn{1}{l|}{512} & \multicolumn{1}{l|}{14} & \multicolumn{1}{l|}{5} & \multicolumn{1}{l|}{12544}  & \multicolumn{1}{l|}{2091} \\ \hline
			\multicolumn{1}{|l|}{CONV12} & \multicolumn{1}{l|}{3} & \multicolumn{1}{l|}{512} & \multicolumn{1}{l|}{512} & \multicolumn{1}{l|}{14} & \multicolumn{1}{l|}{5}  & \multicolumn{1}{l|}{12544}  & \multicolumn{1}{l|}{2091}\\ \hline
			\multicolumn{1}{|l|}{CONV13} & \multicolumn{1}{l|}{3} & \multicolumn{1}{l|}{512} & \multicolumn{1}{l|}{512} & \multicolumn{1}{l|}{14} & \multicolumn{1}{l|}{5}   & \multicolumn{1}{l|}{12544}  & \multicolumn{1}{l|}{2091} \\ \hline
			\multicolumn{8}{l}{	*Note: q=32bit, M=32KB, 1 PE/Router} 
		\end{tabular}}
	\end{center}
	\label{table-INAVgg}
\end{table}
Similarly, for multiple $E$ PEs/router Equation \ref{eq6_roundsINA} can be written as:
\begin{equation} \label{eq6_roundsEINA}
INA_{\#E} =  \Big \lceil \frac{F}{N \cdot E} \cdot \frac{O \times O}{\lfloor \frac{N}{P_\#} \rfloor} \Big \rceil
\end{equation}
It is also to note that, with the increase in computation capacity the performance does not scale linearly since the addition of $E$ PEs/router will increase the communication load in the network as well. 

\subsection{Router Support}
\begin{figure}
	\centering
	\includegraphics[width=0.9\linewidth]{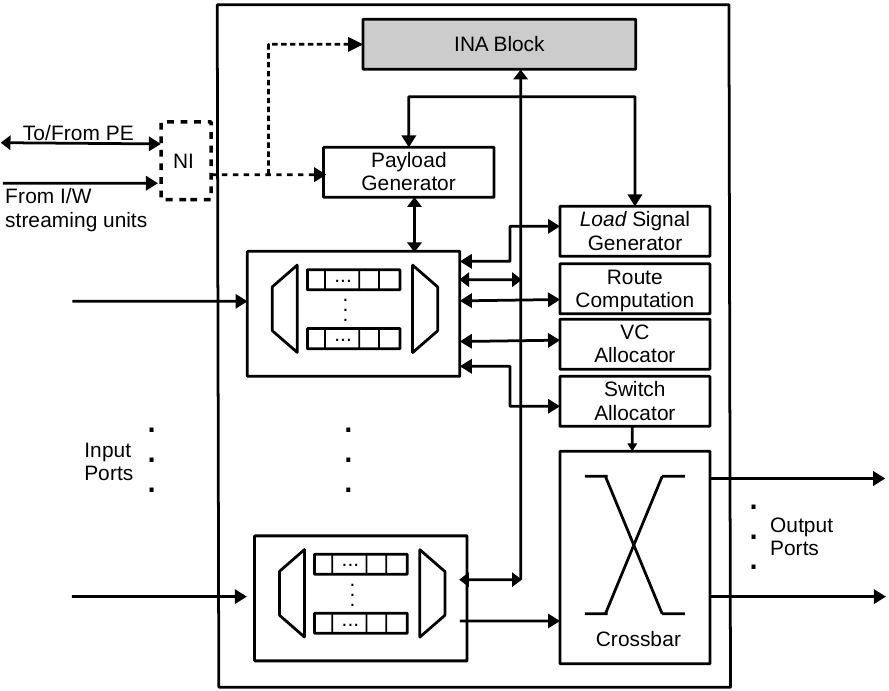}
	\caption{INA Router Microarchitectural}
	\label{fig-ina_router}
\end{figure}

\begin{figure}
	\centering
	\bigskip
	\includegraphics[width=1\linewidth]{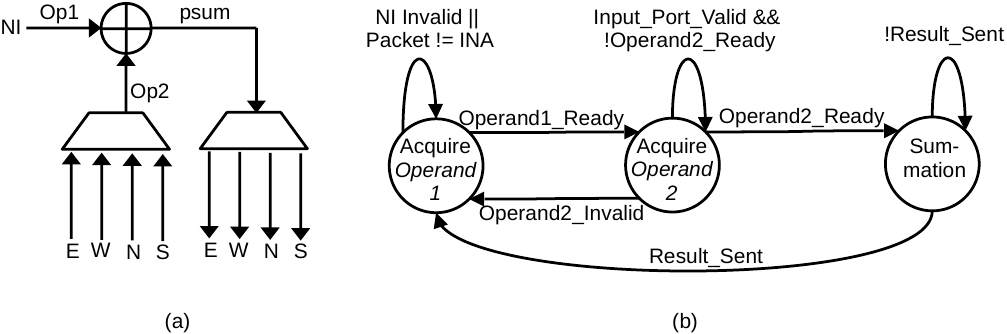}
	\caption{INA Support (a) INA block (b) INA control logic}
	\label{fig-ina_control}
\end{figure}

Fig. \ref{fig-ina_router} shows the changes required to make in the existing router architecture from \cite{jsa} to support the INA. The INA block (Fig. \ref{fig-ina_control}(a)) is added along with the required control (Fig. \ref{fig-ina_control}(b)) and signals to support the INA as shown in Fig. \ref{fig-ina_control}. The INA block is responsible for monitoring the operands from the incoming local port i.e., NI, and from the neighboring node i.e., N, S, E, W port. The INA performs the psum addition without letting the incoming psum exit the router. INA fully utilizes the existing router pipeline and does not affect the existing pipeline stage. The four-stage router pipelines \cite{noc} are sufficient to perform the psum accumulation similar to \cite{jsa} while implementing the gather router.
\par
 Steps \textcolor{black}{ \circled{1}} - \textcolor{black}{ \circled{3}} from Fig. \ref{fig-in_network}(b) can be equivalently mapped to the states $Acquire\ Operand\ 1$, $Acquire\ Operand\ 2$ and $Summation$ states respectively from Fig. \ref{fig-ina_control} (b). The source of $Operand\ 1$ is the incoming packet from the neighboring node which contains a portion of the partial sum. The source of $Operand\ 2$ is the NI of the local PE which calculates the other portion of the partial sum based on the set of weights and input activations distributed to this node and intends to forward it to the next node to complete the partial sum addition.

\par
During the DNN execution, weights are distributed to different nodes in part. The number of parts depends on the memory size of the node and the size of weights for a CONV layer. The scheduler or the top level controller will assign at the runtime i.e., during streaming which node should initiate the INA packet. This is usually the first node that holds the first part of the weight during the distribution.
\par
The purpose of this work is to show the effectiveness of INA in DNN execution by evaluating the network efficiency. There are multiple ways of implementing the adder unit in the router to achieve INA. Various analog adders have been proposed based on ReRAM technology one such example is \cite{reram}. To adopt the ReRAM based structure, we need some additional structures like the digital to analog converter and vice versa. There are other fast digital adders \cite{adder}-\cite{adder3} as an alternative to analog adders. For the multiple PEs/router scenario we can further extend the adders into a simple SIMD/Vector adder unit, some of the alternative choices are presented in \cite{simd_adder}-\cite{vector_adder}. In this work, our primary goal is to show the efficiency of the INA, hence we prefer to use a digital adder from \cite{adder} to validate our concept.

\begin{table}
	\vspace{0.45cm}
	\caption{Network Configuration for INA Simulation}
	\begin{center}
			\resizebox{0.95 \columnwidth}{!}{
		\begin{tabular}{|l|l|}
			\hline
			Topology         & 8x8 Mesh
			\\ \hline
			Virtual Channels & 2
			\\ \hline
			Latency     & router: 4 cycles, link: 1 cycle
			\\ \hline
			Buffer Depth     & 4 flits
			\\ \hline
			Flit Size        & 128 bits/flit
			\\ \hline
			Gather Payload   & 32 bits
			\\ \hline
			Number of PE per router	& 1,2,4,8	
			\\ \hline
			Gather Packet Size      & \begin{tabular}[c]{@{}l@{}}3,5,9 flits/packet for multiple PEs/router\end{tabular}
			\\ \hline
			Unicast Packet Size      & \begin{tabular}[c]{@{}l@{}}2,3 flits/packet  for multiple PEs/router \end{tabular}
			\\ \hline
		\end{tabular}}
	\end{center}
	\label{table-network-ina}
\end{table}

\section{Performance Evaluation} \label{performance-ina}
We assume that there is a higher level entity or a mapping framework similar to  \cite{tpu}, \cite{shidiannao} that does the task of mapping neurons to the PEs, streaming the inputs and weights in parts, controlling timing for better synchronization without stalls, so that our focus is on evaluating the performance of the on-chip network.
\par
To evaluate the performance of INA for the WS dataflow model, we ran simulations for different CNN workloads on 8x8 mesh-based NoCs modified with a two-way streaming architecture \cite{jsa}. For the WS dataflow model, all the weights are streamed in full or in parts as needed depending on the memory availability, these weights are stored in the local memory of the PE and used for multiple rounds of MAC operation for all the input activations. Hence, input activation will not be streamed, and the PE should not start the MAC operation until all the weights are streamed to the local memory of the PE. In this section, we describe the experiment settings, followed by presenting the results.

\subsection{Experiment Setup}
Table \ref{table-network-ina} shows the NoC setting used for performance analysis. To accommodate the gather payload after INA, we have used various gather flit sizes. After the INA, the gather packet can have 1 flit to 9 flits depending upon the number of nodes used to get the accumulation result. For the 1 PE/router case, if the INA is performed on 2 nodes, then the gather packet needs to collect the result of 4 nodes on 8x8 mesh. This information is identified at the compile time by the higher level entity or a mapping framework. 
\par
We compare the WS dataflow model with and without INA both using gather packets in terms of the runtime latency and power consumption, we further extended the result to compare the INA-enabled WS dataflow model with the OS dataflow model with gather support from \cite{jsa}. We assume that the INA is using a similar digital adder proposed in \cite{adder} which is fast and has various bit widths suitable for our analysis, we also assume that the accumulation latency in both cases, with INA and without INA are comparable. We use the parameters obtained from Pytorch framework \cite{pytorch} to model the traces for the NoC. Accumulation happens using INA in the router as explained in Fig. \ref{fig-in_network}. We have used a cycle-accurate C++ based NoC simulator \cite{popnet} to simulate the generated traces for AlexNet \cite{alexnet}, ResNet-50 \cite{resnet50}, and VGG-16 \cite{vgg16}. Orion 3.0 \cite{orion} is used to estimate the power consumption for NoC.

\subsection{Results}
\begin{figure}[ht]
	\centering
	\includegraphics[width=0.98\linewidth]{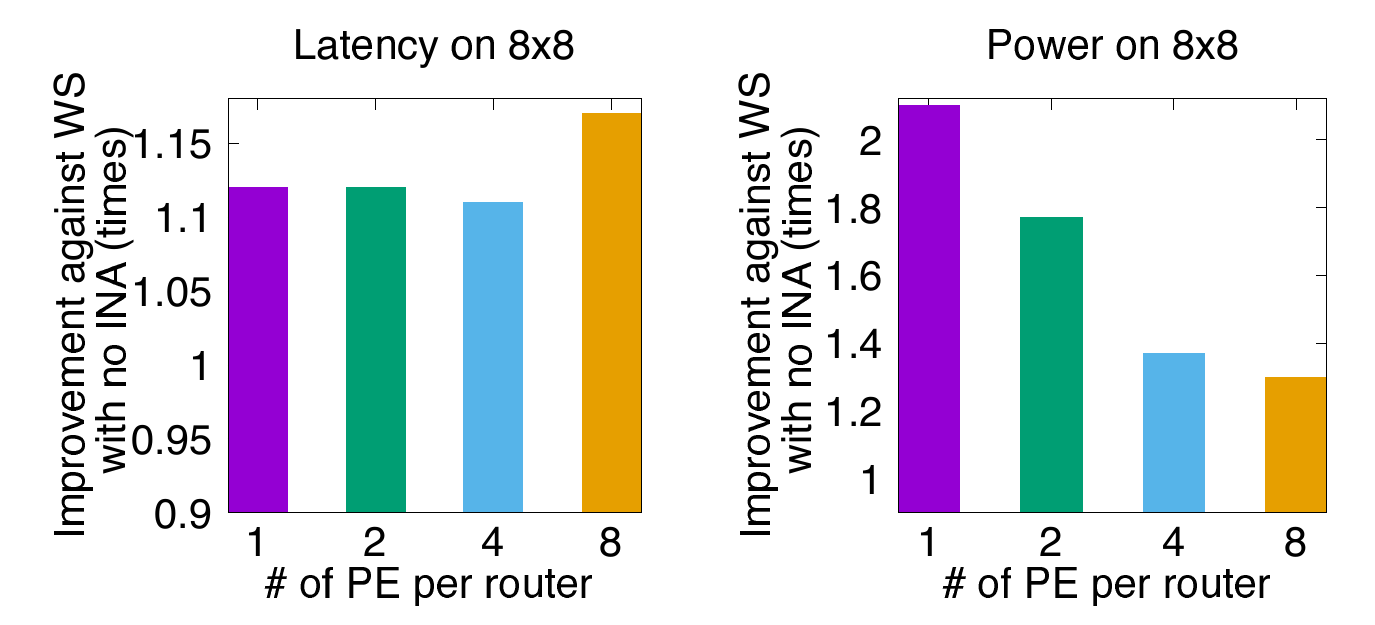}
	\caption{Improvement on total runtime latency (a) and power consumption (b) for AlexNet \cite{alexnet} over WS without INA for different number of PEs/router}
	\label{fig-inaresultAlexnet}
\end{figure}

\begin{figure}[ht]
	\centering
	\includegraphics[width=0.98\linewidth]{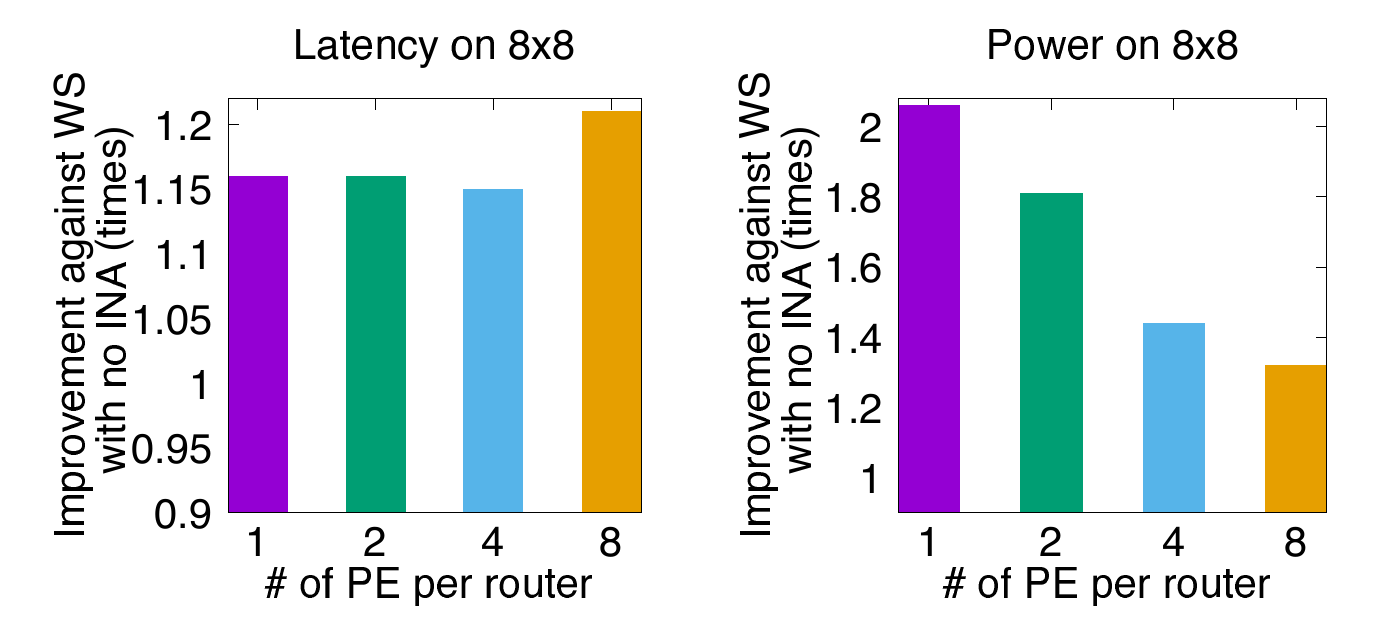}
	\caption{Improvement on total runtime latency (a) and power consumption (b) for ResNet-50 \cite{resnet50} over WS without INA for different number of PEs/router}
	\label{fig-inaresultResnet50}
\end{figure}

\begin{figure}[ht]
	\centering
	\includegraphics[width=0.98\linewidth]{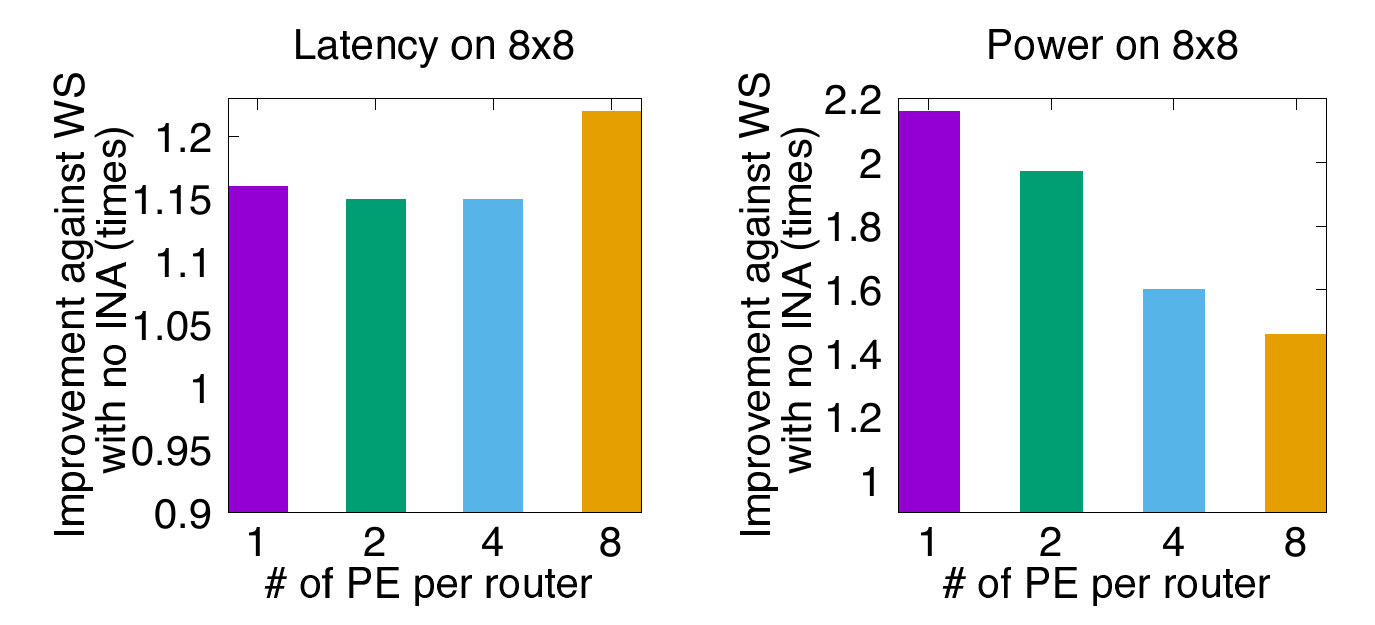}
	\caption{Improvement on total runtime latency (a) and power consumption (b) for VGG-16 \cite{vgg16} over WS without INA for different number of PEs/router}
	\label{fig-inaresultVgg16}
\end{figure}

Fig. \ref{fig-inaresultAlexnet} (a), (b) shows the improvement in the total runtime latency and power consumption of the WS dataflow model with INA against the one without INA case for all convolution layers in AlexNet \cite{alexnet} on 8x8 mesh-based NoCs. We can see that INA can improve the latency up to $1.17\times$ and power consumption up to  $2.1\times$ compared to without INA case. A similar improvement is seen across other DNN workloads as shown in Fig. \ref{fig-inaresultResnet50} and Fig. \ref{fig-inaresultVgg16} for ResNet \cite{resnet50} and VGG-16 \cite{vgg16}, respectively.

\par \bigskip
For all the workloads, the improvement in latency is almost similar in 1, 2, and 4 PEs/router, and the improvement increases for 8 PEs/router. The latency improvement in this comparison is determined by the packet size used for both INA and without INA. For 1, 2, 4 PEs/router, both the cases will use a similar number of flits per packet, and hence we can also see a similar improvement in latency. However, for 8 PEs/router packet size should increase, and without INA case larger packet size adds up to more latency. We can also see that VGG-16 on average has a larger improvement than other workloads due to the multiple rounds of INA ($INA_\#$) needed in VGG-16. Even though ResNet-50 is bigger in terms of the number of CONV layers than VGG-16, most of the ResNet-50 does not need to split the weights among multiple PEs ($I_\#$) and hence ResNet-50 is not showing the highest improvement even with a larger network model.
\par
As for the power improvement, a smaller number of PEs shows the highest improvement. As the number of PEs increases, the number of flits per gather packet should also increase to accommodate all the payloads which contribute to the additional power in the case of INA. Dynamic flit would have made an impact here but in this experiment, we have assumed a static packet size. However, without INA, packets do not need to increase the flit size since the addition is happening between two nodes and a smaller flit can be used to move the partial sum for accumulation. A similar trend for performance is seen across different workloads where VGG-16 is the highest performing due to the similar reasons as explained for the latency improvement.
\par
\begin{figure}[h!]
	\centering
	\includegraphics[width=0.98\linewidth]{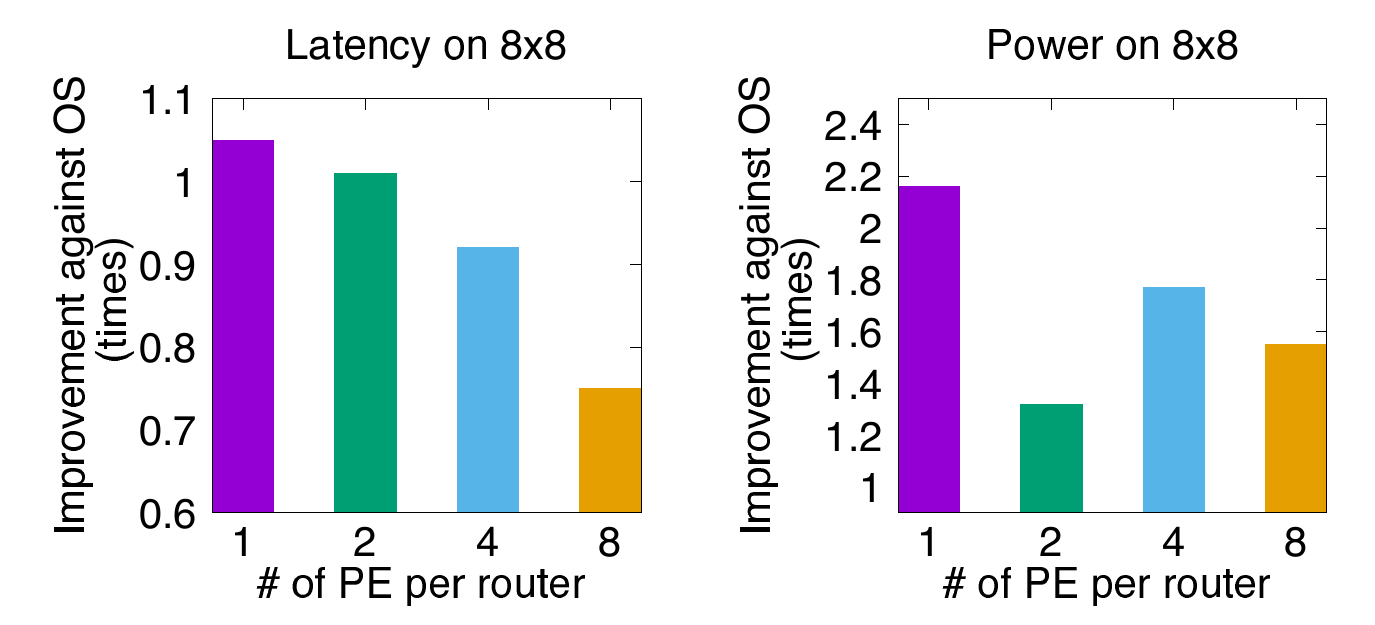}
	\caption{Improvement on total runtime latency (a) and power consumption (b) for AlexNet \cite{alexnet} over OS for different number of PEs/router}
	\label{fig-ws-osresultAlexnet}
\end{figure}
\begin{figure}[h!]
	\centering
	\includegraphics[width=0.98\linewidth]{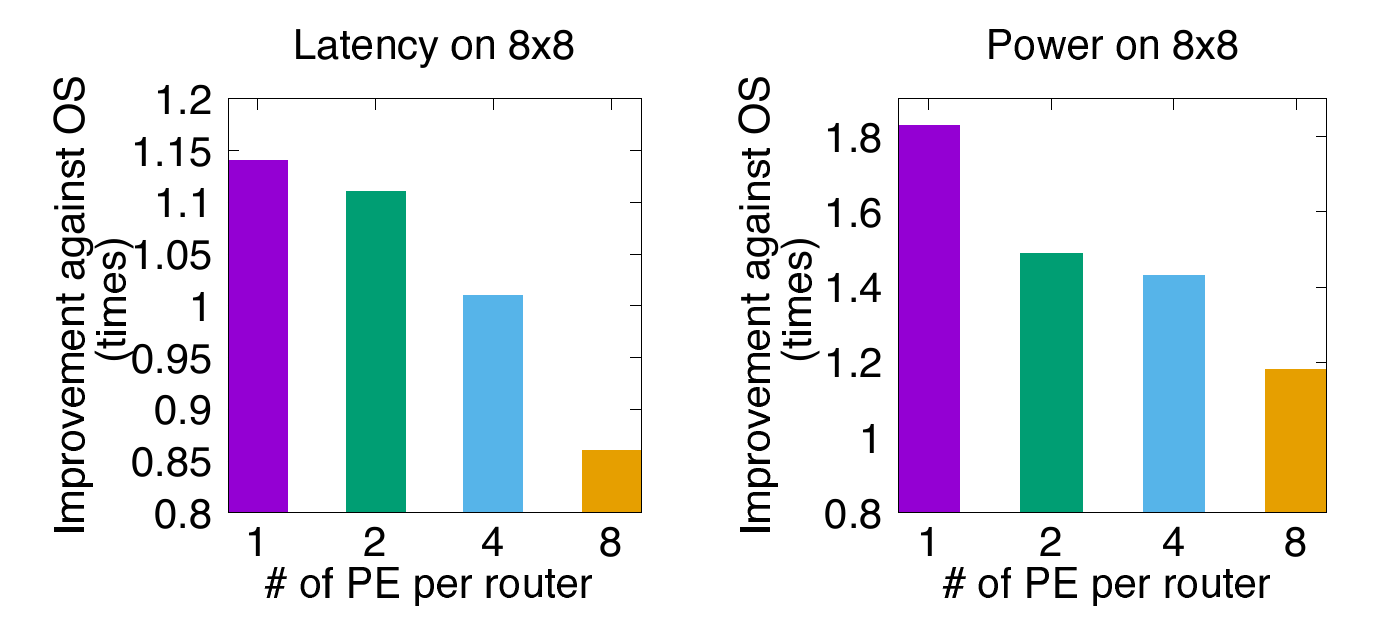}
	\caption{Improvement on total runtime latency (a) and power consumption (b) for ResNet-50 \cite{resnet50} over OS for different number of PEs/router}
	\label{fig-ws-osresultResnet50}
\end{figure}
\begin{figure}[h!]
	\centering
	\includegraphics[width=0.98\linewidth]{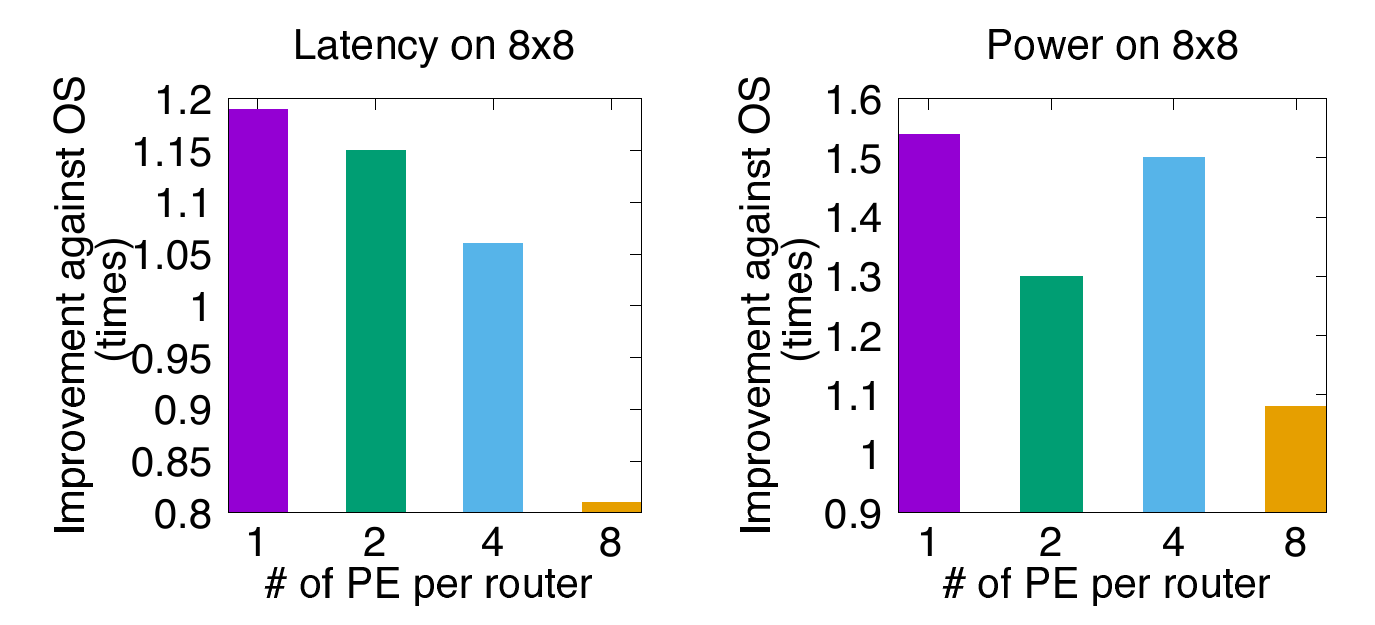}
	\caption{Improvement on total runtime latency (a) and power consumption (b) for VGG-16 \cite{vgg16} over OS for different number of PEs/router}
	\label{fig-ws-osresultVgg16}
\end{figure}

Fig. \ref{fig-ws-osresultAlexnet} - Fig. \ref{fig-ws-osresultVgg16} shows the comparison between the WS dataflow model with INA and gather supported and OS dataflow model with gather supported. The latency result of WS is degrading with the number of PEs increasing because, for the WS dataflow model, weights need to be distributed before the psum accumulation begins. As the number of PEs increases the distribution of weights also takes longer due to the larger packet size. However, on the OS dataflow model both the weight and input distribution happen together and hence the psum accumulation can begin earlier. As for the power improvement, the WS dataflow model outperforms the OS dataflow model in all the cases. This improvement is mainly due to the better reuse of weights leading to less streaming than the OS dataflow model. We see fluctuations in power improvement going from 1 PE/router to 2 PEs/router and from 4 PEs/router to 8 PEs/router because this is the boundary where a change in the flit size happens due to the increase in the number of PEs in the router.

\section{Conclusion} \label{conclusion-ina}
DNN workloads can be executed with a variety of dataflow models. Supporting different dataflow models is important to ensure effective data reuse during the DNN execution. In this paper, we present the In-Network Accumulation (INA) architecture to support partial sum accumulation without ejecting the packet from the network on a WS dataflow model. To support the psum accumulation, an INA block enclosing a digital adder will be added to each router. We performed the simulation on different DNN workloads and showed $1.22\times$ improvement in the runtime latency and $2.16\times$ improvement in the power consumption. Further comparison with the OS dataflow model with gather support, the WS dataflow model with INA achieves up to $1.19\times$ latency improvement and $2.16 \times$ improvement in the power consumption across different DNN workloads. In our future work, this method will be explored further with other dataflow models and NoC topologies.

\section*{Acknowledgment}

This work is supported in part by the National Science Foundation under grant no. 1949585.

\end{document}